\documentclass[reprint,aps,prd,floatfix]{revtex4-2}
\usepackage{graphicx}
\usepackage{bm,amsmath,amssymb}
\newcommand{\pd}[3]{\frac{\partial^{#3} #1}{\partial {#2}^{#3}}} 

\begin{document}
\title{Galaxy clusters in high definition: a dark matter search}

\author{Geoff Beck}
\email[Email: ]{geoffrey.beck@wits.ac.za}

\author{Michael Sarkis}
\affiliation{School of Physics and Centre for Astrophysics, University of the Witwatersrand,
Johannesburg, Wits 2050, South Africa.}

\begin{abstract}
	Recent radio-frequency probes, with the ATCA and ASKAP telescopes, have proven themselves to be at the forefront of placing indirect limits on the properties of dark matter. The latter being able to substantially exceed the constraining power of Fermi-LAT data. However, these observations were based only on dwarf galaxies, where magnetic field uncertainties are large. Here we re-examine the case for galaxy clusters, often ignored due to substantial diffuse radio backgrounds, by considering the extrapolation of known cluster surface brightness profiles down to scales observable with MeerKAT. Despite large baryonic backgrounds, we find that clusters can be competitive with dwarf galaxies. Extrapolated Coma data being able to rule out WIMPs of mass $< 700$ GeV annihilating via $b$-quarks. This is while having lesser uncertainties surrounding the magnetic field and diffusive environment. Such compelling results are possible due to a clash between the inner shape of the dark matter halo and the flat inner profile of radio halos which is most pronounced for NFW-like Einasto profiles, the presence of which having some supporting evidence in the literature. 
\end{abstract}

\maketitle

\section{Introduction}

The nature of Dark Matter (DM) remains a major open question in modern cosmology and particle physics. So-called `indirect' probes of DM, via the consequences of annihilation or decay in cosmic structures, have made major strides in ruling out annihilation models dominated by $b$-quarks and $\tau$-leptons for Weakly Interacting Massive Particles (WIMPs) with masses $\lesssim 100$ GeV through gamma-ray telescopes like Fermi-LAT~\cite{Fermidwarves2016,Hoof_2020}. Recently, radio-frequency probes have begun to realise the potential~\cite{Regis_2014,regis2017,Beck_2019,Cook:2020,Chan:2020wyg,Vollmann:2020a,Basu:2021,Regis_2021} that had previously been argued for~\cite{Colafrancesco2006,gsp2015,Colafrancesco:2015ola,gs2016,Beck_2019}. However, the majority of the radio efforts have focused on dwarf galaxies, as galaxy clusters, despite being heavily DM dominated, tend to host relatively large baryonic background emissions. In Chan et al 2020~\cite{Chan:2020wyg}, the authors look to produce tight constraints on DM in a high redshift galaxy cluster via characterising the cosmic-ray synchrotron contribution. In this work we aim to explore how making use of high angular resolution radio observations of galaxy clusters can contribute to placing powerful limits on DM despite the presence of baryonic backgrounds. This has a significant advantage over the previous method~\cite{Chan:2020wyg}: it is far less uncertain.

The significance of this work is that new, high-resolution, radio observatories like MeerKAT are now online (with the SKA to follow shortly). Thus, exploring their potential in multiple DM dominated environments is a matter of urgency. Currently, the observed trend in the diffuse radio halos of galaxy clusters is that their spatial profile is exponential~\cite{murgia2009}. This presents an opportunity to constrain the properties of DM if the halo density profile follows a non-cored profile, as the clash between the shape of predicted DM emissions and the observed profile at small radii could be highly limiting to potential annihilation cross-sections. It is therefore necessary to consider what the halo shape of clusters tends to be. In Newman et al 2013 and Collet et al 2017~\cite{2013ApJ...765...25N,Collett_2017} the authors' observational results suggest either shallow inner slopes for DM halos or a cored Navarro-Frenk-White (NFW) halo, in contradiction to cold DM simulations that indicate cuspy halos. However, the authors in Mamon et al 2019~\cite{mamon2019b}, find somewhat different results: weak evidence from the WINGS cluster sample ruling out cored NFW profiles in low redshift clusters and no evidence for deviation from plain NFW and NFW-like Einasto profiles. Notably, these results hold down to $0.03 r_{200}$ which, since clusters typically have a concentration parameter $\sim 5$, means that no evidence for deviation from NFW profiles emerges even below the characteristic scale of the halos. In He at al 2020~\cite{He_2020}, the authors also find steeper inner halo slopes from simulated halos than the aforementioned observed values~\cite{2013ApJ...765...25N,Collett_2017}. This is reconciled by noting the difference between asymptotic and mass-weighted mean profile slopes, with the steeper asymptotic values from simulation~\cite{He_2020} having means consistent with shallower values found in observations~\cite{2013ApJ...765...25N,Collett_2017} in the observed regions of the clusters. It seems clear then that there is some evidence in favour of profiles like NFW in galaxy clusters. Notably, the clusters we focus on, being Coma and Ophiuchus, do not display any statistically significant preference between cored and cuspy NFW halos~\cite{coma-halo-2003,durret2015}, suggesting the data is not sufficient to probe substantially below the scale radius of the DM halos. This all means we can have some confidence in using NFW and NFW-like Einasto profiles, while also considering a shallowly cusped profile for good measure.

In this work we find that, with NFW profiles, nearly an order of magnitude improvement in annihilation cross-section upper limits is possible from cluster halo data in the Coma and Ophiuchus clusters when the smallest observable scale is set 10 arcseconds. Since MeerKAT is capable of imaging substantially smaller scales, down to around 5 arcseconds~\cite{Knowles_2022}, we find that high-resolution re-observation of halo-hosting clusters can provide highly competitive limits on WIMP DM, even exceeding a recent ASKAP study of the Large Magellanic Cloud~\cite{Regis_2021} which, under reasonable assumptions about the diffusive environment, rules out WIMPs with masses below 500 GeV annihilating via quarks.

This paper is structured to first review the formalism of radio emissions from WIMP annihilation in section~\ref{sec:emissions}. The galaxy cluster data will then be detailed in section~\ref{sec:sample}. Results are presented in section~\ref{sec:current} and discussed in \ref{sec:disc}.

\section{Radio emissions from dark matter}
\label{sec:emissions}
Radio emissions are produced by DM annihilation when relativistic electrons/positrons (electrons from here on) are products of this process. These electrons are injected continuously into the DM halo over a long period of time. In addition to this, halo environments are commonly magnetised. Thus, we need to consider long-term evolution of the injected electrons before we determine the resulting synchrotron emissions.

\subsection{Diffusion of electrons - Green's functions}
To determine our required electron distributions $\psi$, we must solve an equilibrium form of the diffusion-loss equation:
\begin{equation}
\vec{\nabla}\cdot\left(D(E,\vec{x})\vec{\nabla}\psi\right) + \pd{}{E}{}\left[b(E,\vec{x})\psi\right] + Q_{e}(E,\vec{x}) = 0 \; , \label{eq:difusion-eq}
\end{equation}
where $Q_e$ is the source function from DM injection, $D(E,\vec{x})$ is the diffusion function, and $b(E,\vec{x})$ is the energy loss function. Note that Eq.~(\ref{eq:difusion-eq}) does not prevent super-luminal diffusive motion~\cite{Aloisio2009}. However, this effect only becomes substantial for ultra-high energy cosmic rays~\cite{Aloisio2009}. The source function is given by
\begin{equation}
Q_{e}=\frac{1}{2}\left(\frac{\rho_\chi(\vec{x})}{M_{\chi}}\right)^2{\langle \sigma{V} \rangle} \left. \psi\right\rvert_{\mathrm{inj}} \; , \label{eq:source}
\end{equation}
where $\rho_\chi$ is the DM density, $M_\chi$ is the DM mass, $\langle \sigma{V} \rangle$ is the velocity-averaged annihilation cross-section, and $\left. \psi\right\rvert_{\mathrm{inj}}$ is the spectrum of electrons/positrons injected per DM annihilation. 
To facilitate solution of the equation we will use a Green's function method, which requires the diffusion and loss functions have no spatial dependencies.  
Therefore, we define the diffusion function, under the assumption of Kolmogorov turbulence, via~\cite{Blasi_2001}:
\begin{equation}
D(E) = \ 3\times 10^{28} \left(\frac{d_{0,\mathrm{kpc}}^2 E_{\mathrm{GeV}}}{\overline{B}_{\mu\mathrm{G}}}\right)^{1/3}  \ \mathrm{cm}^2 \ \mathrm{s}^{-1}   \; , \label{eq:diff}
\end{equation}
where $d_0 \sim 15$ kpc is the coherence length of the magnetic field~\cite{bonafede2010}, $d_{0,\mathrm{kpc}} = \left(\frac{d_0}{1 \; \mbox{kpc}}\right)$, $\overline{B}$ is the average magnetic field, $\overline{B}_{\mu\mathrm{G}} = \left(\frac{\overline{B}}{1 \; \mu \mathrm{G}}\right)$, and $E_{\mathrm{GeV}} = \left(\frac{E}{1 \; \mathrm{GeV}}\right)$.
We will also explore a Bohmian diffusion case where~\cite{Blasi_2001}
\begin{equation}
D_{\mathrm{Bohm}}(E) = 3.3 \times 10^{22} \frac{E_{\mathrm{GeV}}}{\overline{B}_{\mu\mathrm{G}}}  \ \mathrm{cm}^2 \ \mathrm{s}^{-1}   \; . \label{eq:diff-bohm}
\end{equation}

The energy-loss function is given by 
\begin{equation}
\begin{aligned}
b(E) & = b_{\mathrm{IC}} E_{\mathrm{GeV}}^2 + b_{\mathrm{sync}} E_{\mathrm{GeV}}^2 B_{\mu\mathrm{G}}^2 \;\\ & + b_{\mathrm{Coul}} \overline{n}_{\mathrm{cm}3} \left(1 + \frac{1}{75}\log\left[\frac{\gamma}{\overline{n}_{\mathrm{cm}3}}\right]\right) \\& + b_{\mathrm{brem}} \overline{n}_{\mathrm{cm}3} E_{\mathrm{GeV}}\;,
\end{aligned}
\label{eq:loss}
\end{equation}
where $\gamma = \frac{E}{m_e c^2}$ with $m_e$ being the electron mass, $\overline{n}$ is the average gas density, and $\overline{n}_{\mathrm{cm}3} = \left(\frac{\overline{n}}{1 \; \mbox{cm}^{-3}}\right)$. The coefficients $b_{\mathrm{IC}}$, $b_{\mathrm{sync}}$, $b_{\mathrm{Coul}}$, $b_{\mathrm{brem}}$ are the energy-loss rates from ICS, synchrotron emission, Coulomb scattering, and bremsstrahlung. These coefficients are given by $0.25\times 10^{-16}(1+z)^4$ (for CMB target photons), $0.0254\times 10^{-16}$, $6.13\times 10^{-16}$, $4.7\times 10^{-16}$ in units of GeV s$^{-1}$. The average quantities $\bar{n}$ and $\bar{B}$ are computed within $r \leq r_s$ ($r_s$ being the characteristic scale of the DM halo), ensuring they accurately reflect the environment of the majority of annihilations. 

The equilibrium solutions to Eq.~(\ref{eq:difusion-eq}) are given by~\cite{baltz1999,baltz2004,Colafrancesco2006}
\begin{equation}
\psi (r,E) = \frac{1}{b(E)}  \int_E^{M_\chi} d E^{\prime} \, G(r,\Delta v) Q_e (r,E^{\prime}) \; ,
\end{equation}
where $G$ is the Green's function, given by:
\begin{align}
G(r,\Delta v) = & \frac{1}{\sqrt{4\pi\Delta v}} \sum_{n=-\infty}^{\infty} (-1)^n \int_0^{r_{\mathrm{max}}} d r^{\prime} \; \frac{r^{\prime}}{r_n} f_{G,n} \; , \\ 
f_{G,n} & =  \left( \mathrm{e}^{-\frac{\left(r^{\prime} - r_n\right)^2}{4\Delta v}} - \mathrm{e}^{-\frac{\left(r^{\prime} + r_n\right)^2}{4\Delta v}} \right)\frac{Q_e(r^{\prime})}{Q_e(r)} \; ,
\end{align}
here
\begin{equation}
\Delta v =  v(u(E)) - v(u(E^{\prime})) \; ,
\end{equation}
with
\begin{equation}
\begin{aligned}
v(u(E)) = & \int_{u_{\mathrm{min}}}^{u(E)} dx \; D(x) \; , \\
u (E) = & \int_E^{E_{\mathrm{max}}} \frac{dx}{b(x)} \; . \\ 
\end{aligned}
\end{equation}


\newcommand{\tr}{\tilde{r}}
\newcommand{\tE}{\tilde{E}}

\subsection{Diffusion of electrons - ADI method}
As an alternative solution approach we implement an Alternating Direction Implicit (ADI) method, in which the diffusion and loss functions keep their full spatial dependence. In this method we set up a multi-dimensional grid over space and energy, and make use of an operator-splitting technique to solve for the equilibrium distribution iteratively. 

It is worth noting, at this point, that although this method is referred to in the literature as ADI, this label is slightly misleading. We will use the ADI term, in keeping with existing literature, but note that this method would be more appropriately referred to as an operator splitting method.

\subsubsection{Diffusion and loss functions}
For this method we replace $\overline{B}$ and $\overline{n}$, in the definition of the loss function $b$, with $B(r)$ and $n(r)$ respectively. The diffusion function we define as
\begin{equation}
D(E) = 10^{29}\left(\frac{B(r)}{B(0)}\right)^{-\frac{1}{3}}E_{\mathrm{GeV}}^{\frac{1}{3}} \ \mathrm{cm}^2 \ \mathrm{s}^{-1}  \; , \label{eq:diff-adi}
\end{equation}
following Regis et al 2017~\cite{regis2017} in their method of including the spatial dependence of the magnetic field. The constant coefficient was chosen to closely match that of Green's function method.

\subsubsection{Crank-Nicolson scheme}
The solution for each dimension in our grid uses a generalised Crank-Nicolson scheme~\cite{press2007},
 which is a method of finite-differencing that uses an average of both explicit and implicit differencing terms. This allows it to gain the stability of an implicit method, while maintaining second-order accuracy. In fact, this method turns out to be unconditionally stable for any time-step size $\Delta t$, which is useful since the diffusion-loss equation considered here contains processes that operate on vastly different time-scales. For arbitrary diffusion and loss functions, we write the general scheme, as in \textcite{regis2015,strong1998},
\begin{multline}
	\dfrac{\psi_i^{n+1}-\psi_i^n}{\Delta t} = \dfrac{\alpha_1 \psi_{i-1}^{n+1} - \alpha_2 \psi_{i}^{n+1} + \alpha_3 \psi_{i+1}^{n+1}}{2\Delta t} \\[0.5ex]
	+ \dfrac{\alpha_1 \psi_{i-1}^{n} - \alpha_2 \psi_{i}^{n} + \alpha_3 \psi_{i+1}^{n}}{ 2\Delta t} + Q_{e,i} \; ,  
\end{multline}
where temporal indices are given by $n$ and dimensional (space, energy) indices by $i$. Each of the $\alpha$ coefficients contain the diffusion and loss functions, and are found by matching this equation to the finite-differenced form of the diffusion-loss equation. 
By isolating the implicit and explicit terms, we obtain the following:
\begin{multline}\label{eqn:cn}
	-\dfrac{\alpha_1}{2}\psi^{n+1}_{i-1} + \left(1+\dfrac{\alpha_2}{2}\right)\psi^{n+1}_{i} - \dfrac{\alpha_3}{2}\psi^{n+1}_{i+1} \\
	= Q_{e,i}\Delta t + \dfrac{\alpha_1}{2}\psi^{n}_{i-1} + \left(1- \dfrac{\alpha_2}{2}\right)\psi^{n}_{i} + \dfrac{\alpha_3}{2}\psi^{n}_{i+1} \; .
\end{multline}
This now represents a system of linear equations in the overall updating equation $A\psi^{n+1} = B\psi^{n} + Q_e$, where $A$ and $B$ are tri-diagonal matrices containing the $\alpha$-coefficients. 

\subsubsection{Operator Splitting}
Since the diffusion-loss equation considered here is 2-dimensional, the Crank-Nicolson scheme as presented above would need to be generalised further, and would thus lose the relative simplicity of the tridiagonal matrix equation. Instead, we make use of an operator splitting technique in which the two dimensions are treated independently, and then solved in alternating steps using the 1-dimensional Crank-Nicolson scheme. This method has been successfully used before, in the public code package \verb|galprop|~\cite{strong1998} and in~\textcite{regis2015}. 

This method is implemented as follows. Firstly, we make the simplifying assumption of spherical symmetry, so that $\vec{x} \rightarrow r$, where $r$ is the radius from the centre of the halo. We then transform the variables $E$ and $r$ to use a logarithmic scale, to better account for the large physical scales involved, \textit{i.e.} $\tE = \log_{10}(E/E_0)$ and $\tr = \log_{10}(r/r_0)$, where $E_0$ and $r_0$ are chosen scale parameters. With these modifications to the diffusion-loss equation, we then find a finite-difference scheme for each of the diffusion and energy loss operators as follows:
\begin{multline}
	\dfrac{1}{r^2}\pd{}{r}{}\left(r^2D\pd{\psi}{r}{}\right) \rightarrow \\
	(r_0\log(10)10^{\tr_i})^{-2}\left[\dfrac{\psi_{i+1}-\psi_{i-1}}{2\Delta \tr}\left.\left(\log(10)D	+ \pd{D}{\tr}{}\right)\right\vert_{i}\right. \\
    + \left.\dfrac{\psi_{i+1}-2\psi_{i}+ \psi_{i-1}}{\Delta \tr^2}D\vert_{i} \right] \; ,
\end{multline} 
for radius and 
\begin{equation}
	\pd{}{E}{}(b\psi) \rightarrow (E_0\log(10)10^{\tE_j})^{-1}\left[\dfrac{b_j\psi_{j+1}-b_j\psi_{j}}{\Delta \tE}\right] \; ,
\end{equation}
for energy, where $\Delta \tr$ and $\Delta \tE$ represent the radial and energy grid spacings, respectively. Note that in the case of energy losses, we only consider upstream differencing. This corresponds to each grid point's energy loss only depending on those points with equal or higher energies (or, to only $j$ and $j+1$ terms entering into the updating equation).

If these schemes are represented by $\Psi$, we can summarise the solution method with the following steps:
\begin{align}
    \psi^{n+1/2} &= \Psi_{\tE}(\psi^{n}) \; , \\
    \psi^{n+1} &= \Psi_{\tr}(\psi^{n+1/2}) \; .
\end{align}
These steps are then computed on each iteration of the algorithm in turn, updating the value of $\psi$ until convergence is reached and the equilibrium distribution is found.

The forms of $\Psi_{\tE}$ and $\Psi_{\tr}$ are also used to find the values of the $\alpha$-coefficients present in the matrices $A$ and $B$. By equating coefficients with the general 1-dimensional Crank-Nicolson scheme given above, these values are then given by
\begin{equation}\label{eqn:alpha_r}
	\Psi_{\tr}:
	\begin{cases}
		\dfrac{\alpha_1}{\Delta t} &= C_{\tr}^{-2}\left.\left(-\dfrac{\ln(10)D+\frac{\partial D}{\partial\tr}}{2\Delta\tr} + \dfrac{D}{\Delta\tr^2}\right)\right\vert_{i}\;,\\[3ex]
		\dfrac{\alpha_2}{\Delta t} &= C_{\tr}^{-2}\left.\left(\dfrac{2D}{\Delta\tr^2}\right)\right\vert_{i} \; ,\\[3ex]
		\dfrac{\alpha_3}{\Delta t} &= C_{\tr}^{-2}\left.\left(\dfrac{\ln(10)D+\frac{\partial D}{\partial\tr}}{2\Delta\tr} + \dfrac{D}{\Delta\tr^2}\right)\right\vert_{i} \; ,
	\end{cases}
\end{equation}
for the spatial dimension and 
\begin{equation}\label{eqn:alpha_E}
	\Psi_{\tE}:
	\begin{cases}
		\dfrac{\alpha_1}{\Delta t} &= 0 \; ,\\[3ex]
		\dfrac{\alpha_2}{\Delta t} &= C_{\tE}^{-1}\dfrac{b_j}{\Delta E}\; ,\\[3ex]
		\dfrac{\alpha_3}{\Delta t} &= C_{\tE}^{-1}\dfrac{b_{j+1}}{\Delta E} \; ,
	\end{cases}
\end{equation}
for the energy. Here $C_{\tr}=(r_0\log(10)10^{\tr_i})$ and $C_{\tE} = (E_0\log(10)10^{\tE_j})$. 

\subsubsection{Initial and Boundary Conditions}
The initial condition on $\psi$ is simply set as $0$ everywhere. We then use the following Dirichlet and Neumann boundary conditions:
\begin{align}
	\psi &= 0 \; , \qquad \tr = \tr_{\mathrm{max}} \; ,\\
	\pd{\psi}{\tr}{} &= 0 \; , \qquad \tr = \tr_{\mathrm{min}} \; .
\end{align}
The Dirichlet condition is enforced by setting all values of $\psi$ at $\tr=\tr_{\mathrm{max}}$ to 0 at each step during the solution. The Neumann condition is enforced by using alternate $\alpha$-coefficients at the point $\tr=\tr_{\mathrm{min}}$. The substitution of this condition into the diffusion-loss equation leads to a new differencing scheme, with $\alpha$-coefficients given by:
\begin{equation}
	\Psi_{\tr=\tr_{\mathrm{min}}}:
	\begin{cases}
		\dfrac{\alpha_1}{\Delta t} &= 0 \; ,\\[3ex]
		\dfrac{\alpha_2}{\Delta t} &= C_{\tr}^{-2}\left.\left(\dfrac{4D}{\Delta\tr^2}\right)\right\vert_{i} \; ,\\[3ex]
		\dfrac{\alpha_3}{\Delta t} &= C_{\tr}^{-2}\left.\left(\dfrac{4D}{\Delta\tr^2}\right)\right\vert_{i} \; .
	\end{cases}
\end{equation}

\subsubsection{Physical Scales}
The changing of $\psi$ over time has several physical time-scales associated with it. These factors play a role in the numerical stability of the algorithm, and are used when determining its point of convergence. The diffusion and energy loss time-scales (respectively) are calculated as
\begin{equation}\label{eqn:loss_ts}
	\tau_{\mathrm{loss}} = \dfrac{\tE}{b(\tE,\tr)} \; ,
\end{equation}
and
\begin{equation}\label{eqn:diff_ts}
	\tau_{D} = \dfrac{\tr_{\mathrm{min}}^2}{D(\tE,\tr)} \; .
\end{equation}
We estimate the time-scale of changes to $\psi$ using the form $\vert\psi/\frac{\partial\psi}{\partial t}\vert$ and a simple Forward-Time finite difference, resulting in
\begin{equation}\label{eqn:psi_ts}
	\tau_{\psi} = \frac{\psi}{\left\vert\left(\dfrac{\psi^{n+1} - \psi^n}{\Delta t}\right)^{-1}\right\vert} \; .
\end{equation}


\subsubsection{Stability and Convergence}\label{sec:stability}
Since the relevant physical time-scales of energy-loss and diffusion can vary by several orders of magnitude, if the time-steps $\Delta t$ used to update $\psi$ are too small then the number of iterations required to reach the final solution could become extremely large. Otherwise, if the $\Delta t$ are too large, details on smaller scales could lose accuracy. Therefore, as in~\cite{regis2015,strong1998}, we adopt an `accelerated' method of determining the time-steps $\Delta t$. This method involves choosing an initial time-step that is large compared to the time-scales of all physical effects, such that $\Delta t_i \geq \max{\{\tau_{\mathrm{loss}}, \tau_D\}}$, and then running the algorithm with this value until stability is achieved. After this point, we reduce the value of $\Delta t$ and repeat the process until it reaches a value that is lower than any of the physical time-scales. This switching of time-steps allows us to update the electron distribution with all of the relevant time-scales, while saving on computational resources that would be wasted if $\Delta t$ was too small or too large. 

The continuation or termination of this algorithm is managed by several conditions that relate to how the electron distribution changes from one iteration to the next. The final equilibrium solution is only found once the convergence conditions are met, and the convergence conditions are only evaluated when the algorithm is stable. For stability, we ensure that a minimum number of iterations occur for each time-step value in the accelerated method (typically 100 iterations per $\Delta t$). Once the algorithm reaches the lowest value of $\Delta t$ and the minimum number of iterations are complete, we then check for convergence by computing $\tau_{\psi}$ and requiring this to be larger than all relevant physical effect time-scales, \textit{i.e.} $\tau_{\psi} > \max{\{\tau_{\mathrm{loss}},\tau_D\}}$. We also monitor the relative change of $\psi$ between iterations and require that this be lower than some tolerance, typically $\sim 10^{-5}$. Once these conditions are met, any changes to $\psi$ due to energy-losses or diffusion should be negligible and so we consider $\psi$ to be the equilibrium distribution. 


\subsection{Synchrotron emission}
With the equilibrium electron distributions in hand we can determine the resulting synchrotron emissions.
We start with the power emitted by an electron with energy $E$, at frequency $\nu$, and position $r$~\cite{longair1994}:
\begin{equation}
P_{\mathrm{sync}} (\nu,E,r) = \int_0^\pi d\theta \, \frac{\sin{\theta}^2}{2}2\pi \sqrt{3} r_e m_e c \nu_g  F_{\mathrm{sync}}\left(\frac{\kappa}{\sin{\theta}}\right)  \; ,
\label{eq:power}
\end{equation}
where $r_e = \frac{e^2}{m_e c^2}$ is the classical electron radius, $e$ is the electronic charge, $\nu_g = \frac{e B}{2\pi m_e c}$ is the non-relativistic gyro-frequency, and $B$ is the magnetic field strength.
Additionally,
\begin{equation}
\kappa = \frac{2\nu}{3\nu_g \gamma^2}\left[1 +\left(\frac{\gamma \nu_p}{\nu}\right)^2\right]^{\frac{3}{2}} \; ,
\end{equation}
where $\gamma = \frac{E}{m_e c^2}$ and $\nu_p \propto \sqrt{n_e}$ is the plasma frequency. Then the kernel function is given by
\begin{equation}
F_{\mathrm{sync}}(x) \simeq 1.25 x^{\frac{1}{3}} \mbox{e}^{-x} \left(648 + x^2\right)^{\frac{1}{12}} \; .
\end{equation}
To determine the emissions from DM annihilation products we define an emissivity
\begin{equation}
j_{\mathrm{sync}} (\nu,r) = \int_{0}^{M_\chi} dE \, 2 \psi(E,r) P_{\mathrm{sync}} (\nu,E,r) \; ,
\label{eq:emm}
\end{equation}
where $2\psi$ is the approximate sum of electron and positron equilibrium distributions.
The flux, integrated from the target centre out to radius $R$, is then found via
\begin{equation}
S_{\mathrm{sync}} (\nu,R) = \int_0^R d^3r^{\prime} \, \frac{j_{\mathrm{sync}}(\nu,r^{\prime})}{4 \pi (d_L^2+\left(r^{\prime}\right)^2)} \; .
\label{eq:flux}
\end{equation}
Finally, the surface brightness, at a distance $R$ from the target centre, is given by
\begin{equation}
I_{\mathrm{sync}} (\nu,R) = \int dl \, \frac{j_{\mathrm{sync}}(\nu,\sqrt{R^2+l^2})}{4 \pi} \; , 
\label{eq:sb}
\end{equation}
where $l$ is the line-of-sight coordinate and the integral runs over the line of sight through the target at $R$.

\section{Galaxy cluster sample}
\label{sec:sample}
We will be considering two galaxy clusters in this work: Ophiuchus and Coma. We will detail their relevant properties as well as the observed diffuse synchrotron emissions in these targets. 

For both clusters we will make use of the Hernquist-Zhao~\cite{gnfw1,gnfw2} profile:
\begin{equation}
\rho_{\chi} (r) = \frac{\rho_s}{\left(\frac{r}{r_s}\right)^{\alpha_z}\left(1+\frac{r}{r_s}\right)^{3-\alpha_z}} \; ,
\end{equation}
where $\alpha_z$ is the profile index, $\rho_s$, and $r_s$ are the characteristic density and radius respectively. We will consider both $\alpha_z = 1$ (NFW/cusped~\cite{nfw1996}) and $\alpha_z=0.5$ (shallow cusp). In addition, we make use of the Einasto profile~\cite{einasto1968}
\begin{equation}
	\rho_{\chi} (r) = \rho_{s} \exp\left[-\frac{2}{\alpha_e} \left(\left[\frac{r}{r_s}\right]^{\alpha_e} - 1\right)\right]\; ,
\end{equation}
in our case $\alpha_e = 0.17$ to closely mimic the NFW profile in the outer regions of the halo.

For both studied clusters we use a $\beta$-profile for the gas density and a magnetic field profile that follows the gas distribution
\begin{align}
n_e (r) & = n_0 \left(1 + \left(\frac{r}{r_e}\right)^2\right)^{3\beta/2} \; , \\
B(r) & = B_0\left(\frac{n_e(r)}{n_0}\right)^{\eta} \; ,
\end{align}
where $n_0$ is the central gas density, $B_0$ is the central field strength, and $r_e$ is the scale radius of the gas distribution.

\subsection{The Coma cluster}
\begin{figure}[ht!]
	\resizebox{0.8\hsize}{!}{\includegraphics{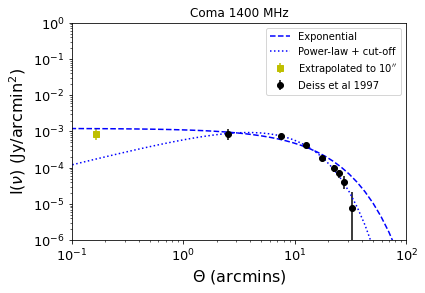}}
	\resizebox{0.8\hsize}{!}{\includegraphics{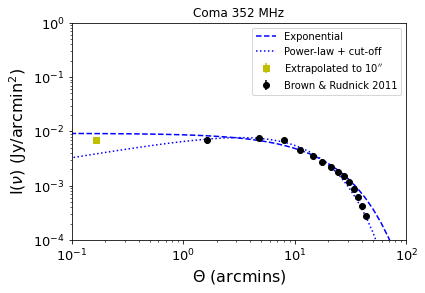}}
	\caption{Radio data sets and extrapolations for the Coma cluster.}
	\label{fig:coma-data}
\end{figure}

For the Coma cluster we make use of two sets of surface-brightness data, from Deiss et al 1997~\cite{deiss1997} at 1.4 GHz as well as Brown \& Rudnick 2011~\cite{brown-rudnick-2011} at 352 MHz. These data sets are pictured in Fig.~\ref{fig:coma-data}. We display the data points themselves, as well as fitting functions and an extrapolated data point. It is notable that the power-law and cut-off model fits both data sets better than the usual exponential model for cluster surface brightnesses~\cite{murgia2009} (though it is unclear what this would imply physically). However, we make use of the more conservative approach when determining any extrapolation of the distribution to smaller angular radii. This we do by replicating the data point at $\sim 2.5$ arcminutes, which agrees well with the exponential fit. 

The values that specify the various relevant properties of Coma are listed in Table~\ref{tab:coma}
\begin{table}[ht!]
	\begin{tabular}{|c|c|c|}
		\hline
		Property & Value & Reference \\
		\hline
		$z$ & 0.0231 & \cite{coma-halo-2003} \\
		$B_0$ & 4.7 $\mu$G & \cite{bonafede2010}\\
		$\eta$ & 0.5 & \cite{bonafede2010}\\
		$n_0$ & $3.49\times 10^{-3}$ cm$^{-3}$ & \cite{chen-clusters-2007}\\
		$\beta$ & -0.654 & \cite{chen-clusters-2007}\\
		$r_e$ & 253 kpc & \cite{chen-clusters-2007}\\
		$M_{\mathrm{vir}}$ & $1.2 \times 10^{15}$ M$_{\odot}$ & \cite{coma-halo-2003} \\
		$R_{\mathrm{vir}}$ &  2.7 Mpc & \cite{coma-halo-2003}\\
		$c_{\mathrm{vir}}$ & 9.4 & \cite{coma-halo-2003}\\ 
		\hline
	\end{tabular}
\caption{Coma cluster properties}
\label{tab:coma}
\end{table}

\subsection{The Ophiuchus cluster}
\begin{figure}[ht!]
	\resizebox{0.8\hsize}{!}{\includegraphics{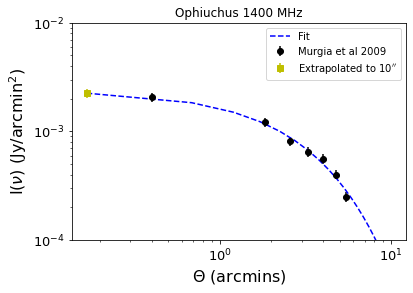}}
	\resizebox{0.8\hsize}{!}{\includegraphics{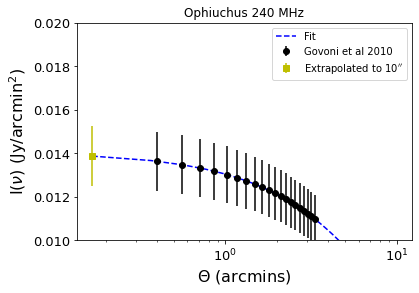}}
	\caption{Radio data sets and extrapolations for the Ophiuchus cluster. The curve labelled fit refers to a fitted exponential function.}
	\label{fig:ophiuchus-data}
\end{figure}

In the case of Ophiuchus, we make use of 1.4 GHz data from Murgia et al 2009~\cite{murgia2009} as well as 240 MHz data from Govoni et al 2010~\cite{govoni2010}. These are displayed in Fig.~\ref{fig:ophiuchus-data}, in a similar manner to those of Coma. Note that we follow Zandanel et al 2013~\cite{zandanel2013} in ascribing a 10\% relative error to these data sets.

The values that specify the various relevant properties of Ophiuchus are listed in Table~\ref{tab:ophiuchus}
\begin{table}[ht!]
	\begin{tabular}{|c|c|c|}
		\hline
		Property & Value & Reference \\
		\hline
		$z$ & 0.028 & \cite{durret2015}\\
		$B_0$ & 10 $\mu$G & \cite{zandanel2013}\\
		$\eta$ & 0.4 & \cite{zandanel2013}\\
		$n_0$ & $6.8\times 10^{-3}$ cm$^{-3}$ & \cite{chen-clusters-2007}\\
		$\beta$ & -0.747 & \cite{chen-clusters-2007}\\
		$r_e$ & 199 kpc & \cite{chen-clusters-2007}\\
		$M_{\mathrm{vir}}$ & $1.1 \times 10^{15}$ M$_{\odot}$ & \cite{durret2015} \\
		$R_{\mathrm{vir}}$ &  2.1 Mpc & \cite{durret2015}\\
		$c_{\mathrm{vir}}$ & 3.0 & \cite{durret2015}\\ 
		\hline
	\end{tabular}
	\caption{Ophiuchus cluster properties}
	\label{tab:ophiuchus}
\end{table}

\section{Dark matter limits from Coma and Ophiuchus}
\label{sec:current}
Here we will begin to display results. These are derived by comparing the data sets specified in Section~\ref{sec:sample} to predicted DM surface-brightnesses and integrated fluxes found according to Section~\ref{sec:emissions} (following the Green's function methodology). This comparison is used to determine 95\% confidence interval upper limits on the DM annihilation cross-section $\langle \sigma V\rangle$. In the process, we consider three annihilation channels: $b\bar{b}$, $\mu^+\mu^-$, and $\tau^+\tau^-$. These correspond to processes like $\chi \chi \to b \bar{b} \to e^+ e^-$.  When plotting our result, the 3 halo profile choices are represented by shaded bands around the solid line (NFW). The upper edge of the band corresponds to the shallow cusp, whereas the lower edge represents the NFW-like Einasto case.

\subsection{The Coma cluster}

\begin{figure}[ht!]
	\resizebox{0.8\hsize}{!}{\includegraphics{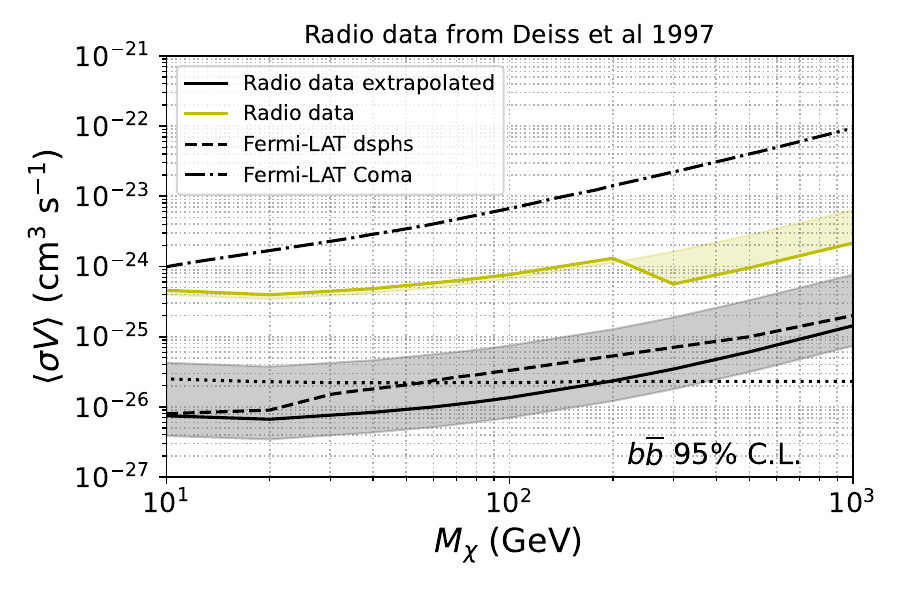}}
	\resizebox{0.8\hsize}{!}{\includegraphics{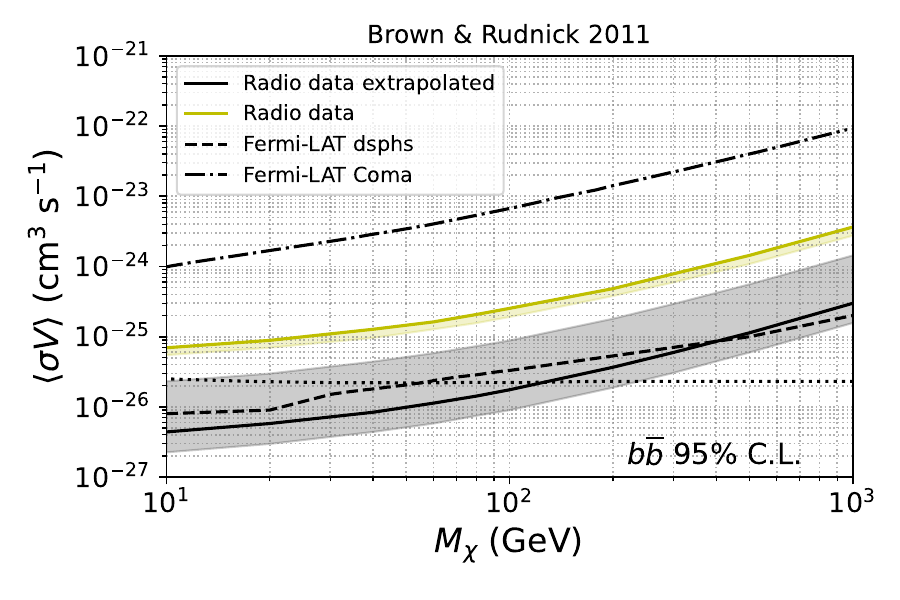}}
	\caption{95\% Confidence interval upper limits on $\langle \sigma V\rangle$, with annihilation via $b$-quarks, derived from the Coma cluster via surface brightnesses. The shaded bands represent variation with halo profile choice. \textit{Top}: Using data set from Deiss et al 1997~\cite{deiss1997}. \textit{Bottom}: Using data set from Brown \& Rudnick 2011~\cite{brown-rudnick-2011}.}
	\label{fig:coma-bb}
\end{figure}

\begin{figure}[ht!]
	\resizebox{0.8\hsize}{!}{\includegraphics{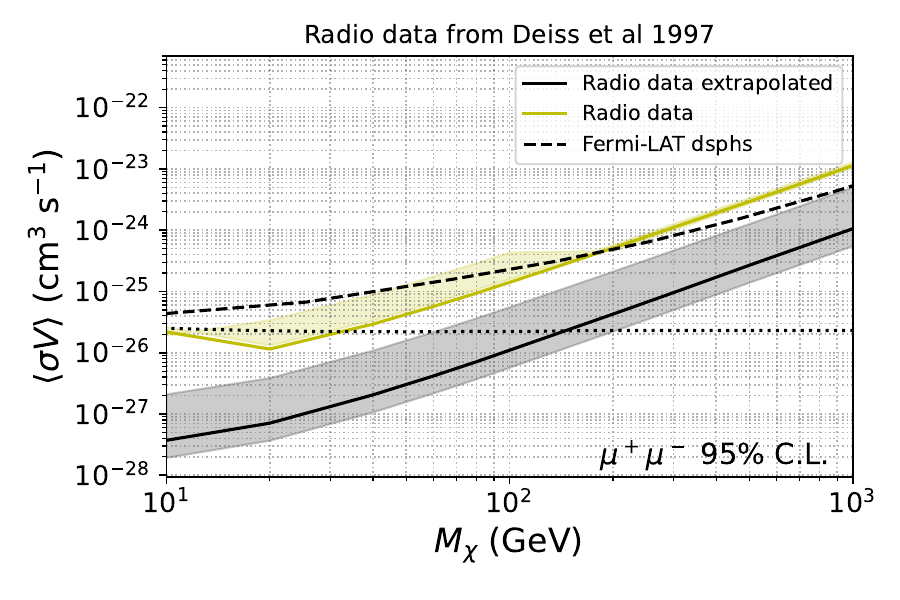}}
	\resizebox{0.8\hsize}{!}{\includegraphics{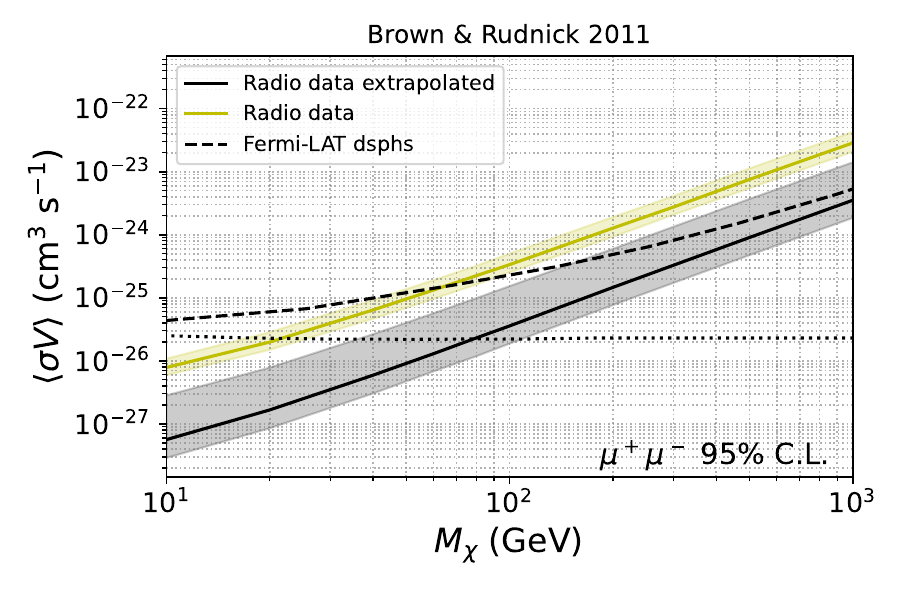}}
	\caption{95\% Confidence interval upper limits on $\langle \sigma V\rangle$, with annihilation via muons, derived from the Coma cluster via surface brightnesses. The shaded bands represent variation with halo profile choice. \textit{Top}: Using data set from Deiss et al 1997~\cite{deiss1997}. \textit{Bottom}: Using data set from Brown \& Rudnick 2011~\cite{brown-rudnick-2011}.}
	\label{fig:coma-mumu}
\end{figure}

\begin{figure}[ht!]
	\resizebox{0.8\hsize}{!}{\includegraphics{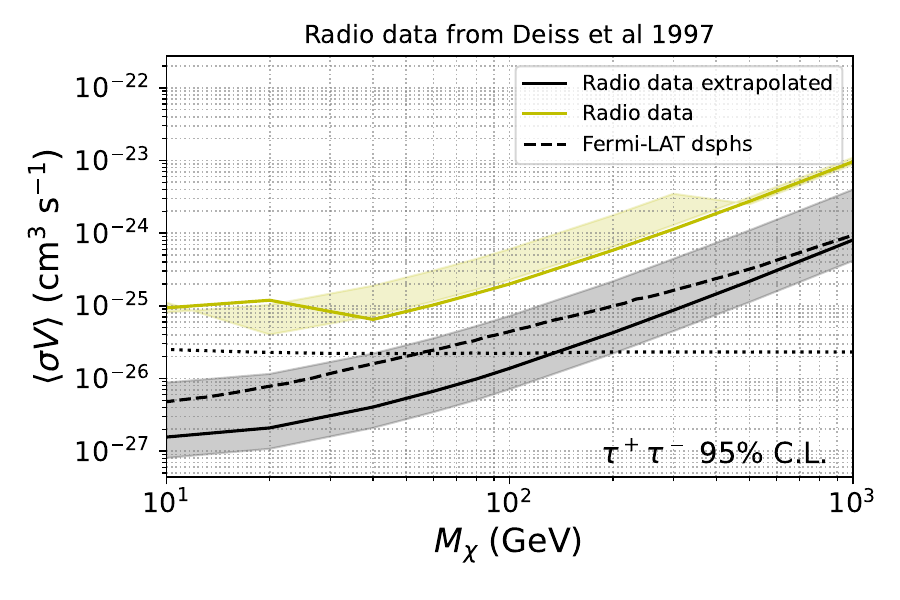}}
	\resizebox{0.8\hsize}{!}{\includegraphics{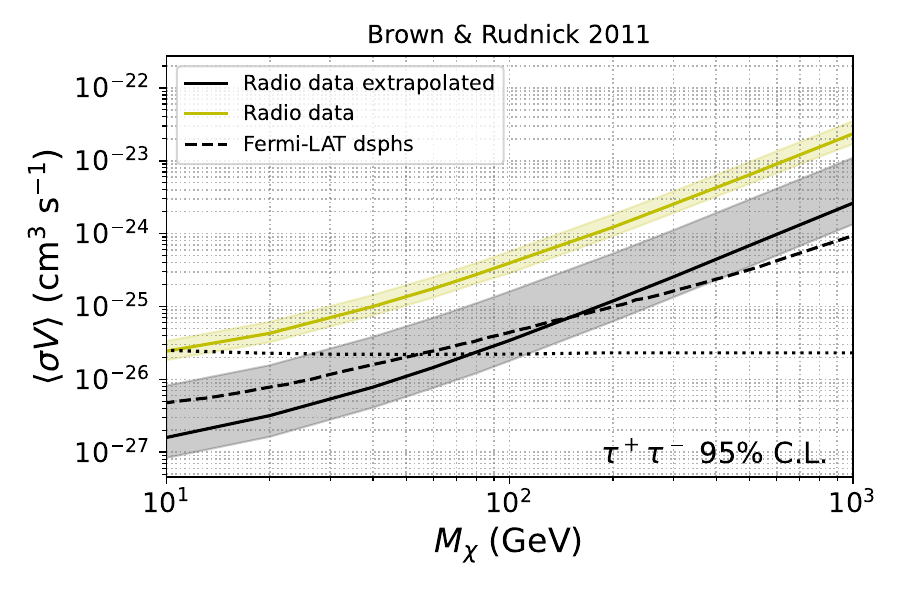}}
	\caption{95\% Confidence interval upper limits on $\langle \sigma V\rangle$, with annihilation via $\tau$-leptons, derived from the Coma cluster via surface brightnesses. The shaded bands represent variation with halo profile choice. \textit{Top}: Using data set from Deiss et al 1997~\cite{deiss1997}. \textit{Bottom}: Using data set from Brown \& Rudnick 2011~\cite{brown-rudnick-2011}.}
	\label{fig:coma-tautau}
\end{figure}

In Figures~\ref{fig:coma-bb} to \ref{fig:coma-tautau} we display the 95\% confidence interval upper limits on the DM annihilation cross-section for our three annihilation channels. For comparison we include the results from the Fermi-LAT dwarf galaxy search~\cite{Albert_2017} and a search in nearby galaxy clusters~\cite{Thorpe_Morgan_2021}. When using the un-extrapolated data, the only channel that out-performs Fermi is annihilation via light leptons for $M_\chi < 200$ GeV. The results are otherwise not competitive, despite the large DM mass within the clusters, as the observed diffuse emission is relatively bright. The lower frequency results displayed from Brown \& Rudnick 2011~\cite{brown-rudnick-2011} are superior in the un-extrapolated cases. However, when we consider extrapolation of the surface brightness profiles down to 10 arcseconds, motivated by the resolution of telescopes like MeerKAT. It is evident that, for both NFW (solid line) and Einasto (bottom of shaded band) halos, there is a dramatic improvement in the potential DM limits. This is due simply to the clash between the flat surface-brightness profile and the cuspy DM density at small halo radii. It should be noted the Einasto case offers twice as good limits as the cusped NFW profile. This is a consequence of the Einasto halo having a higher density near the transition between NFW power-law components. In order to match the Einasto case, the NFW profile requires an extrapolation of the surface brightness data down to scales of $1^{\prime\prime}$ (the former profile's improvement saturates at $\approx 10^{\prime\prime}$ when $\left[\frac{r}{r_s}\right]^{\alpha_e} \approx 0.5$). It is notable that the limits exceed those from Fermi-LAT dwarf-galaxy searches significantly for all three channels across the mass range. In particular, the $b\bar{b}$ channel NFW (Einasto) halo limits reach below the relic cross-section when $M_\chi \lesssim 200 (400)$ GeV. For the shallowly cusped halo (top of shaded band), the limits improve at low masses, but not sufficiently to make the $b$-quark channel competitive with Fermi-LAT. Interestingly, the $b$-quark results are superior to the Fermi-LAT gamma-ray limits from Coma itself~\cite{Thorpe_Morgan_2021} even without extrapolation. 

It is noteworthy that the variation between halo profiles is much larger for the extrapolated cases. This is simply due to the fact that this scenario explores regions of the halo that begin to differ substantially in their predicted surface brightness from DM annihilation. This means that the potential improvment in DM limits will be contingent on higher precision probes of the DM distribution in target clusters. The slightly odd behaviour of the un-extrapolated limits with the Deiss et al 1997~\cite{deiss1997} data is simply a consequence of changing WIMP annihilation spectra with mass creating features due to the distribution of the data and its attendant error bars (particularly the data point at the largest angular radius). 

\subsubsection{Comparison to integrated fluxes}
A common approach in the literature is to rely on integrated fluxes, rather than surface-brightness profiles. Therefore, it is necessary that we contrast the results obtained above with this approach. To do this we will use the results from Thierbach et al 2002~\cite{thierbach2002} and thus limit our integration radius to $\sim 30$ arcminutes.

An important aspect of the integrated fluxes is the effect of DM substructures within the parent halo. Synchrotron emissions cannot benefit from the full flux enhancement as sub-halos are common on periphery of the parent~\cite{jiang2017}, where the magnetic field is weaker. Thus, we will construct the full boost, within the virial radius, according to Sanchez-Conde et al 2017~\cite{sanchez-conde2017} and label this $f_{\mathrm{boost}}$. We will then account for spatial distribution of the substructure and magnetic field via 
\begin{equation}
	\mathcal{B}(R) = 4\pi \int_0^R dr \ r^2 f_{\mathrm{boost}}\left(\frac{B(r)}{B_0}\right) \tilde{\rho}_{\mathrm{sub}}(r) \; ,
\end{equation}
where $R$ is the radius of flux integration and $\tilde{\rho}_{\mathrm{sub}}$ is the sub-halo mass density, from Jiang \& van den Bosch 2017~\cite{jiang2017}, normalised to 1 in the range $r= 0$ to $r = R_{\mathrm{vir}}$. The factor $\mathcal{B}(R)$ will then be multiplied with the parent halo flux to determine the total flux. We scale with $B$, rather than $B^2$, as our cluster fluxes all exhibit such a dependency, likely due to the effect of energy losses. If it turns out that $\mathcal{B}(R) < 1$ we will take $\mathcal{B}(R) = 1$, to represent that there is no significant contribution from sub-halos. To illustrate the effect we note that the Coma cluster would have $f_{\mathrm{boost}} \approx 55$, but has $\mathcal{B} \approx 7.4$ within a radius of 30 arcminutes. 

\begin{figure}[ht!]
	\resizebox{0.8\hsize}{!}{\includegraphics{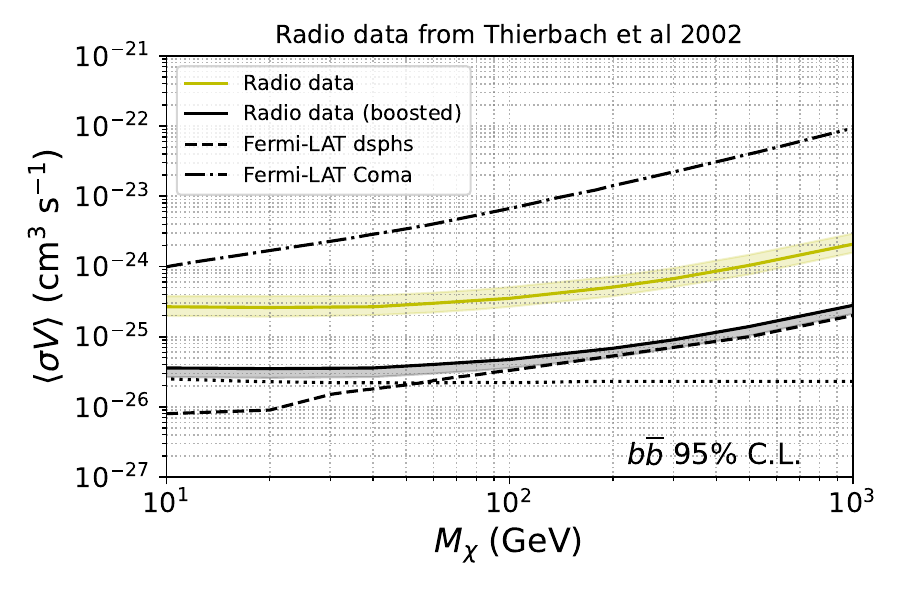}}
	\resizebox{0.8\hsize}{!}{\includegraphics{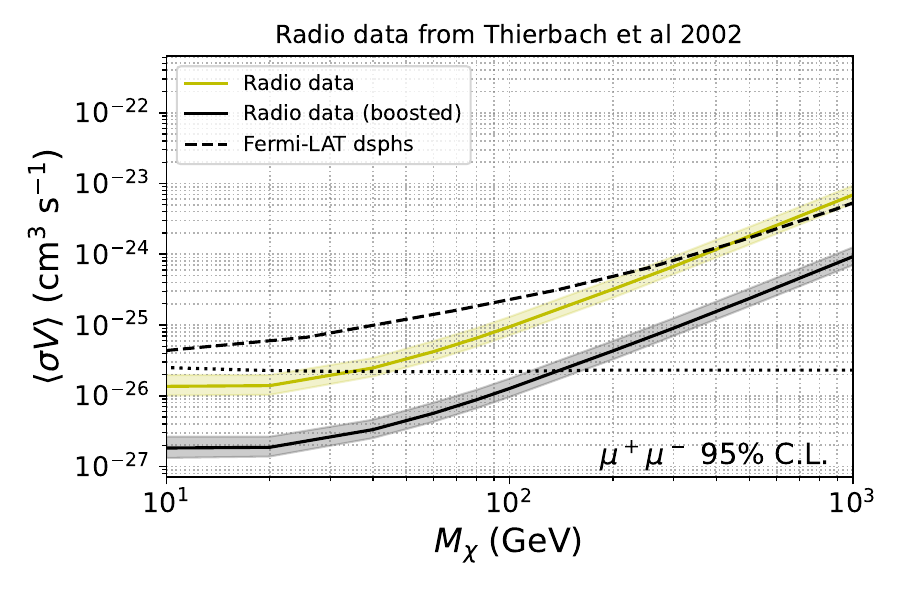}}
	\resizebox{0.8\hsize}{!}{\includegraphics{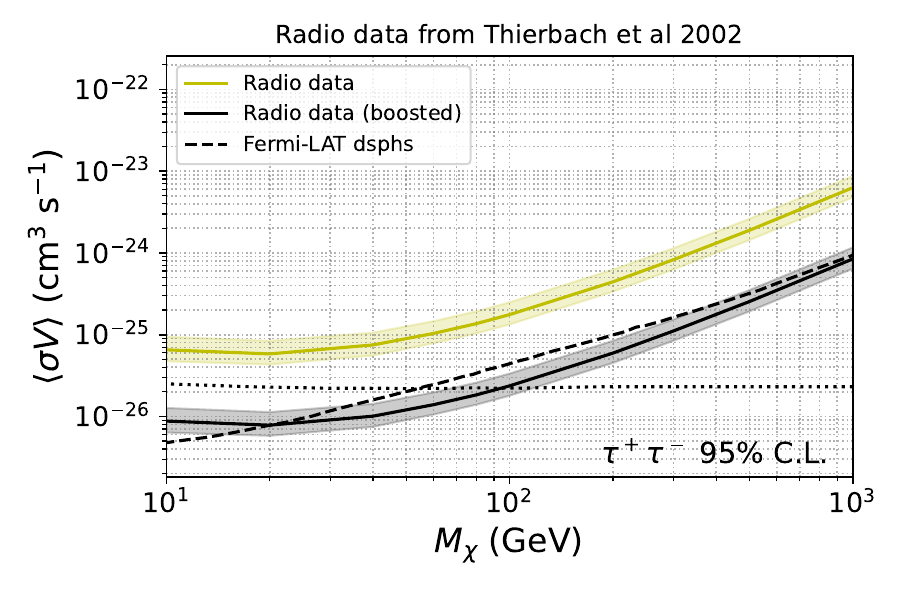}}
	\caption{95\% Confidence interval upper limits on $\langle \sigma V\rangle$ derived from the Coma cluster via integrated flux from Thierbach et al 2002~\cite{thierbach2002}. The shaded bands represent variation due to choice of halo profile. \textit{Top}: annihilation via $b$-quarks. \textit{Middle}: muon channel. \textit{Bottom}: $\tau$-lepton channel.}
	\label{fig:coma-thierbach}
\end{figure}

What is notable about the results displayed in Figure~\ref{fig:coma-thierbach} is that all three halo choices (shaded bands) have very similar results, and the relative behaviour of the $b$-quark and leptonic channel limits is reversed from our previous results. In the surface brightness cases the $b$-quark provides thermal relic cross-section exclusions at larger masses than the other channels. Whereas, for integrated fluxes, this channel is barely excluded at the relic level at all. Boosting is necessary to provide competitive exclusions for the $\tau$ channel. The leptonic channels are inferior to the Einasto surface brightness results but compete well in the case of the NFW profile. However, the necessity of boosting mean that these results should be regarded with caution, due to the uncertainty in predictions of substructure effects. Notably, the magnitude of the boost is the main reason for the differences to previous work with integrated fluxes in Coma~\cite{gs2016}. 

\subsection{The Ophiuchus cluster}

\begin{figure}[ht!]
	\resizebox{0.8\hsize}{!}{\includegraphics{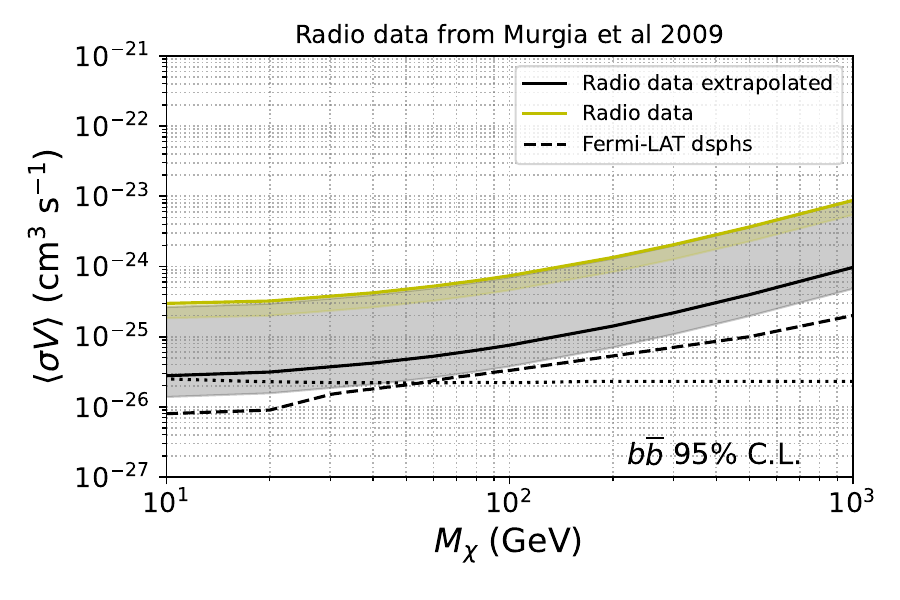}}
	\resizebox{0.8\hsize}{!}{\includegraphics{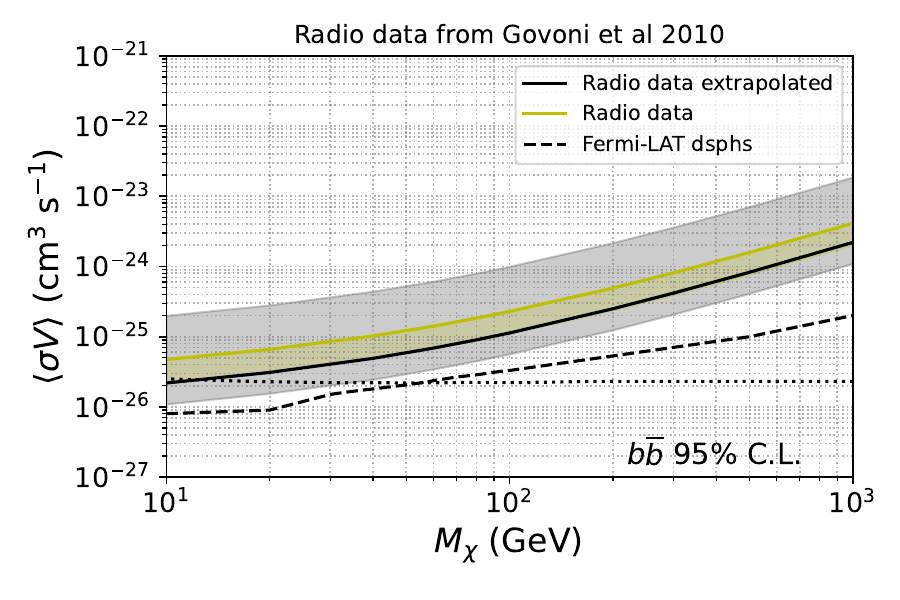}}
	\caption{95\% Confidence interval upper limits on $\langle \sigma V\rangle$, with annihilation via $b$-quarks, derived from the Ophiuchus cluster via surface brightnesses. The shaded bands represent variation with halo profile choice. \textit{Top}: Using data set from Murgia et al 2009~\cite{murgia2009}. \textit{Bottom}: Using data set from Govoni et al 2010~\cite{govoni2010}.}
	\label{fig:oph-bb}
\end{figure}

\begin{figure}[ht!]
	\resizebox{0.8\hsize}{!}{\includegraphics{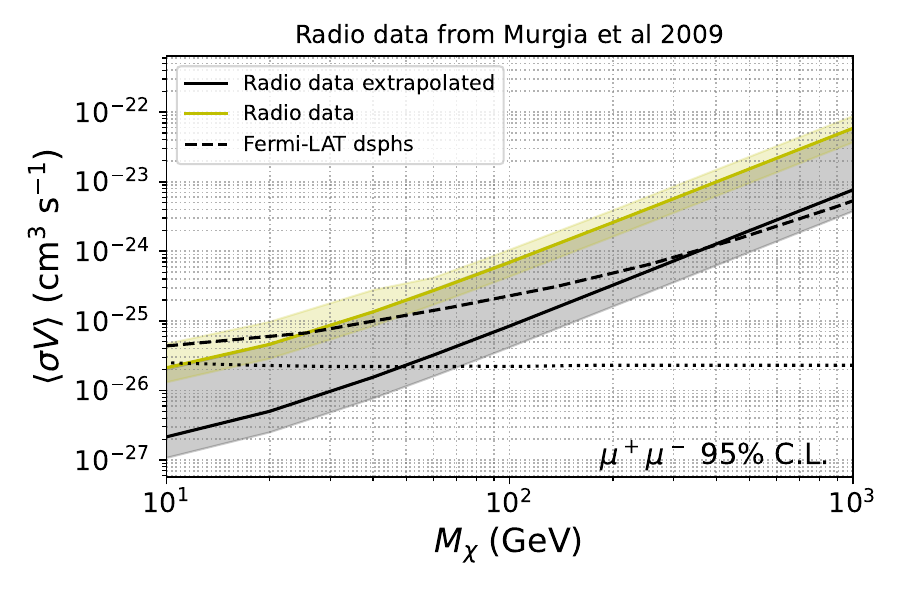}}
	\resizebox{0.8\hsize}{!}{\includegraphics{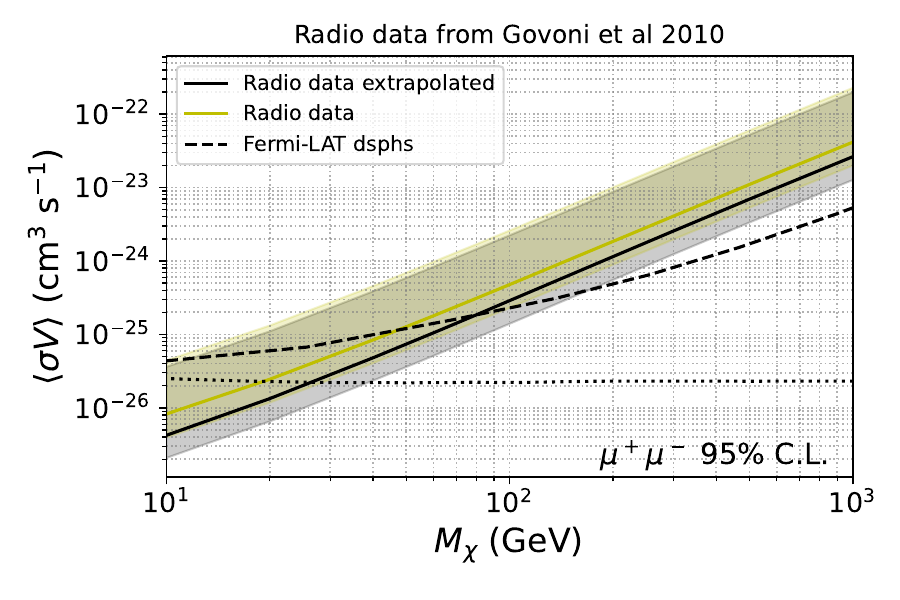}}
	\caption{95\% Confidence interval upper limits on $\langle \sigma V\rangle$, with annihilation via muons, derived from the Ophiuchus cluster via surface brightnesses. The shaded bands represent variation with halo profile choice. \textit{Top}: Using data set from Murgia et al 2009~\cite{murgia2009}. \textit{Bottom}: Using data set from Govoni et al 2010~\cite{govoni2010}.}
	\label{fig:oph-mumu}
\end{figure}

\begin{figure}[ht!]
	\resizebox{0.8\hsize}{!}{\includegraphics{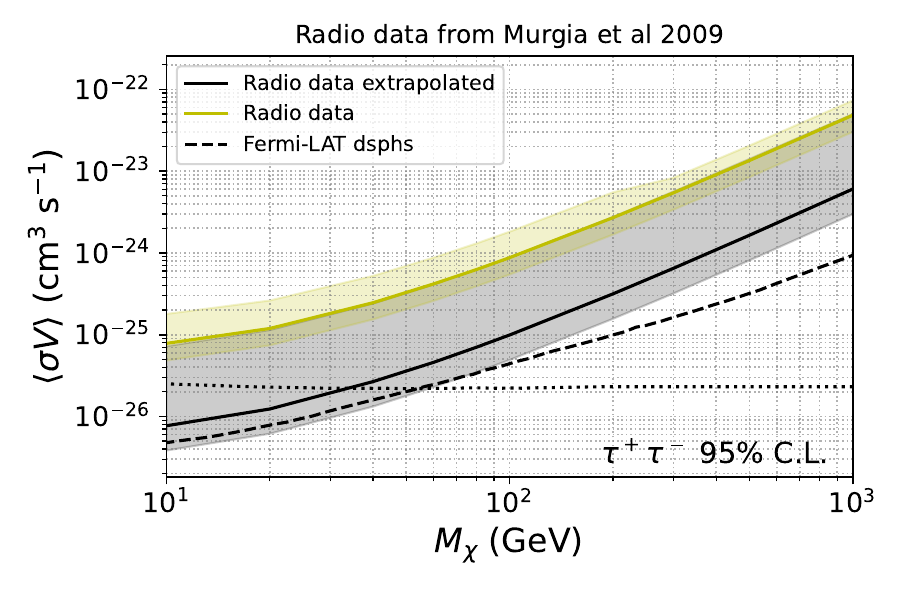}}
	\resizebox{0.8\hsize}{!}{\includegraphics{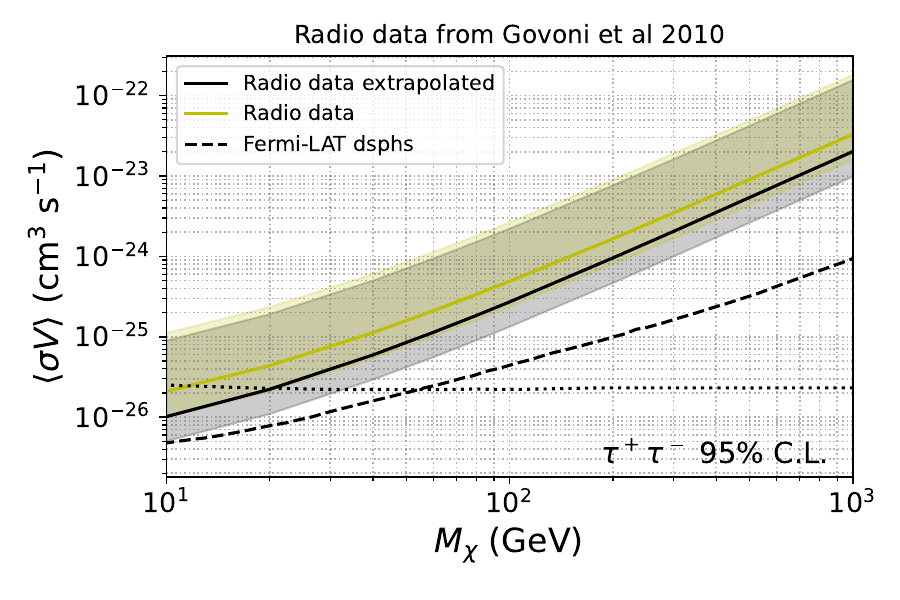}}
	\caption{95\% Confidence interval upper limits on $\langle \sigma V\rangle$, with annihilation via $\tau$-leptons, derived from the Ophiuchus cluster via surface brightnesses. The shaded bands represent variation with halo profile choice. \textit{Top}: Using data set from Murgia et al 2009~\cite{murgia2009}. \textit{Bottom}: Using data set from Govoni et al 2010~\cite{govoni2010}.}
	\label{fig:oph-tautau}
\end{figure}

Our second surface-brightness case is that of Ophiuchus. The results for each annihilation channel are shown in Figs.~\ref{fig:oph-bb} to \ref{fig:oph-tautau}. The un-extrapolated results are somewhat weaker than Coma, with even the light lepton channel struggling to compete with Fermi-LAT. However, at the lower frequency of 240 MHz, we see that the NFW (solid line) and Einasto (bottom of shaded band) halos produce better limits, exceeding those of the un-extrapolated Coma cluster data (especially for lower mass DM). This is due to the larger magnetic field strength in Ophiuchus following Zandanel et al 2013~\cite{zandanel2013}. The shallow cusp case (top of shaded band) does not benefit as strongly from the magnetic field as the halo profile is not so steep in the region of the strongest magnetic field. 
For the extrapolated data, the situation in the Ophiuchus cluster is similar to Coma, in that the extrapolated data point greatly improves limits with NFW and Einasto halos at both frequencies. However, despite the extrapolation making these limits at least competitive with Fermi-LAT in all channels, they are weaker then Coma. Part of this is that the extrapolation is substantially less effective on the low-frequency data, as it already had a data point at $\sim 24$ arcseconds. Similar to Coma, the shallow cusp profile benefits very mildly from the extrapolation and the lower frequency results remain weaker. The differences between Coma and Ophiuchus come down to the surface brightness profile for Coma flattening at much larger scales than in Ophiuchus. It is interesting to note that the $b$-quark channel appears to benefit from this most strongly. The principle difference between this and the leptonic channels being that annihilations yield relatively more high-energy electrons, it seems that this annihilation channel results in somewhat brighter synchrotron emission at small halo radii, an effect which becomes much more significant at larger WIMP masses (at low mass this channel is fainter at all scales).

\subsection{The effect of diffusion parameters}
\begin{figure}[ht!]
	\resizebox{0.8\hsize}{!}{\includegraphics{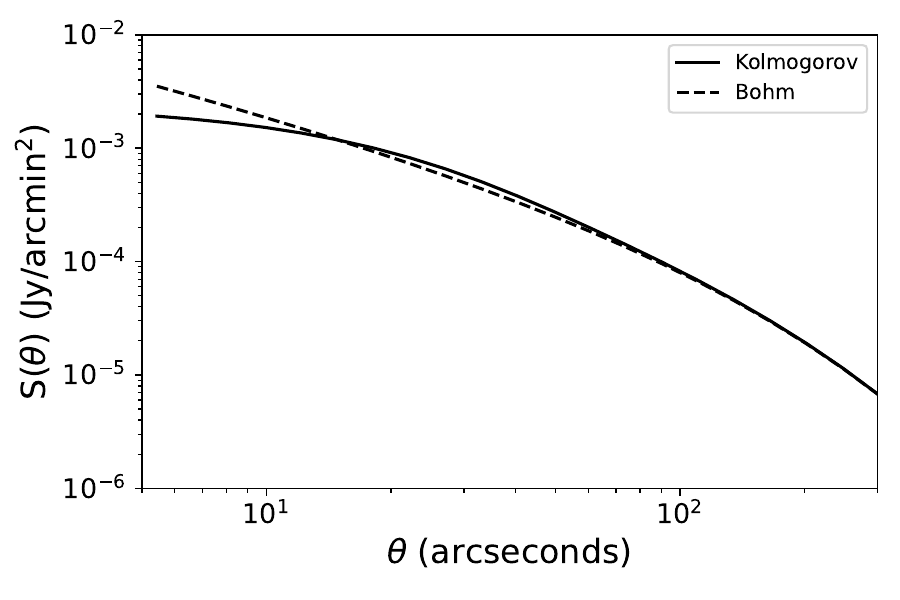}}
	\caption{Effect of diffusion parameters on Coma cluster surface brightnesses for 100 GeV WIMP annihilating via $b$-quarks with $\langle \sigma V\rangle = 10^{-26}$ cm$^3$ s$^{-1}$.}
	\label{fig:diff-compare}
\end{figure}

An important question, when considering the robustness of DM limits derived via radio data, is that of the choice of diffusion parameters. To explore this we consider two sets of assumptions as previously detailed: Kolmogorov and Bohmian diffusion following Blasi 2001~\cite{Blasi_2001}. We display the effect in the Coma cluster using an NFW halo, and 100 GeV WIMP, at 1400 MHz in Figure~\ref{fig:diff-compare}. It is evident that the extremely different diffusion scenarios result in a relatively small difference in surface brightness at small radii only. The maximum difference for $\theta > 10^{\prime\prime}$ is negligible for all studied frequencies and WIMP masses. 

\subsection{ADI method comparison}

\begin{figure}[ht!]
	\resizebox{0.8\hsize}{!}{\includegraphics{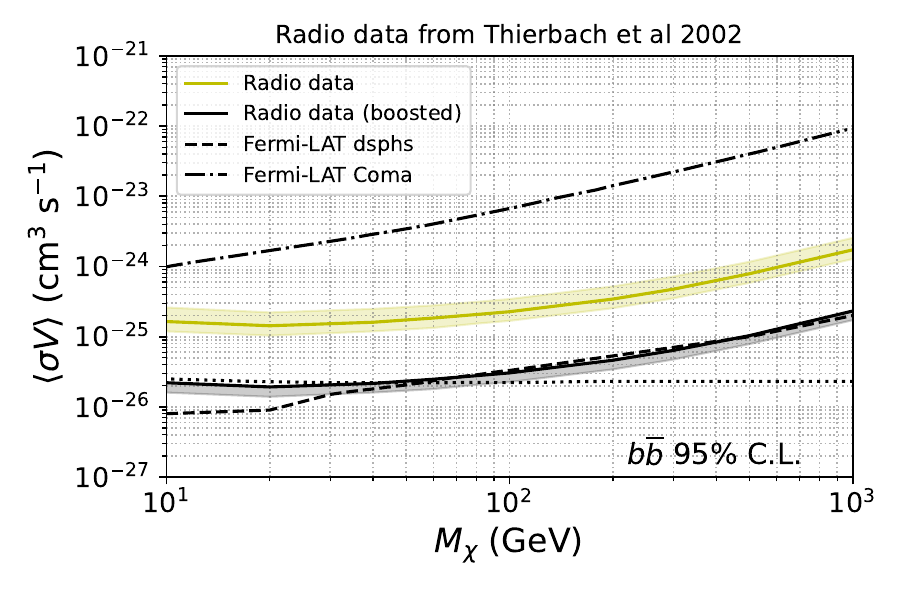}}
	\resizebox{0.8\hsize}{!}{\includegraphics{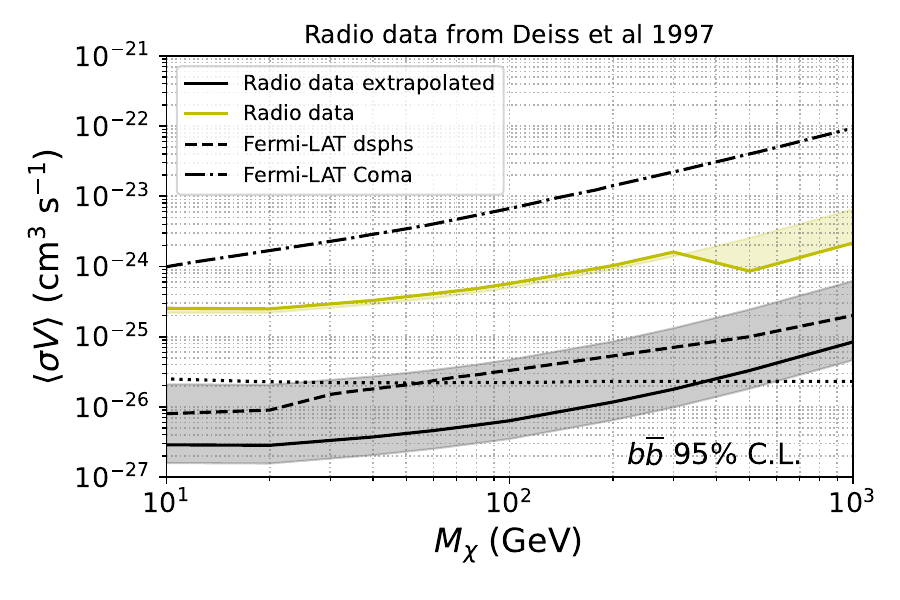}}
	\caption{95\% Confidence interval upper limits on $\langle \sigma V\rangle$, with annihilation via $b$-quarks, derived from the Coma cluster. The shaded bands represent the variation due to choice of halo profile. \textit{Top}: Using data set from Thierbach et al 2002~\cite{thierbach2002}. \textit{Bottom}: Deiss et al 1997~\cite{deiss1997}.}
	\label{fig:adi-bb}
\end{figure}

\begin{figure}[ht!]
	\resizebox{0.8\hsize}{!}{\includegraphics{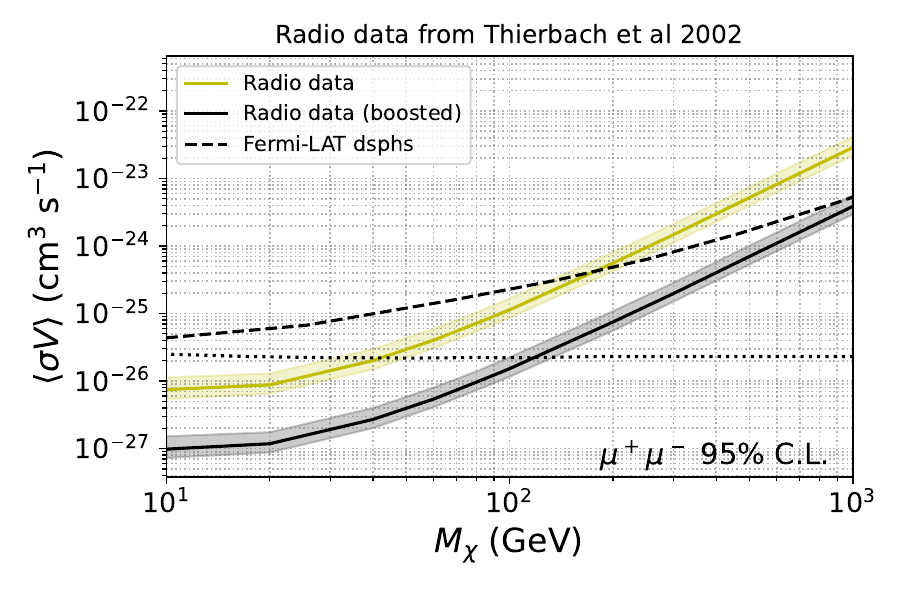}}
	\resizebox{0.8\hsize}{!}{\includegraphics{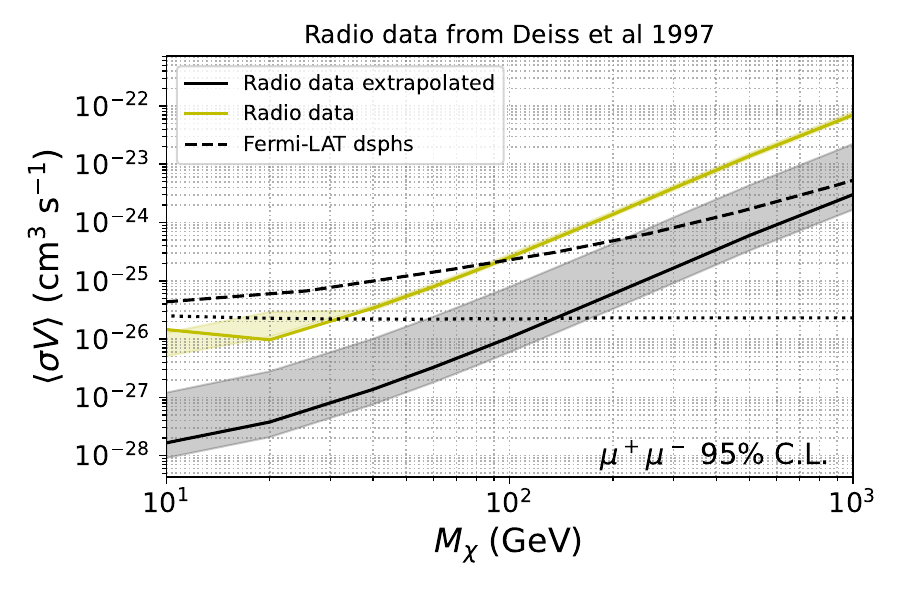}}
	\caption{95\% Confidence interval upper limits on $\langle \sigma V\rangle$, with annihilation via muons, derived from the Coma cluster. The shaded bands represent the variation due to choice of halo profile. \textit{Top}: Using data set from Thierbach et al 2002~\cite{thierbach2002}. \textit{Bottom}: Deiss et al 1997~\cite{deiss1997}.}
	\label{fig:adi-mumu}
\end{figure}

\begin{figure}[ht!]
	\resizebox{0.8\hsize}{!}{\includegraphics{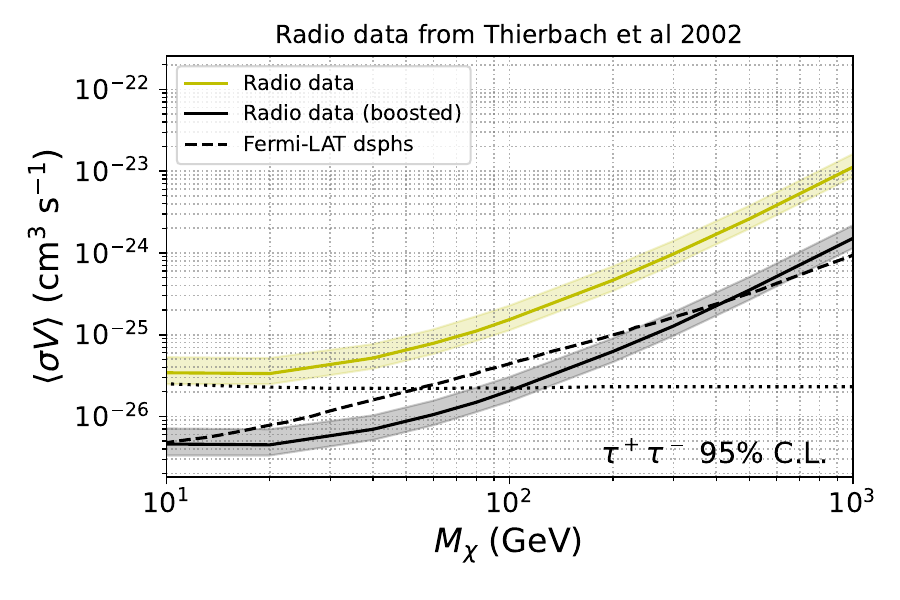}}
	\resizebox{0.8\hsize}{!}{\includegraphics{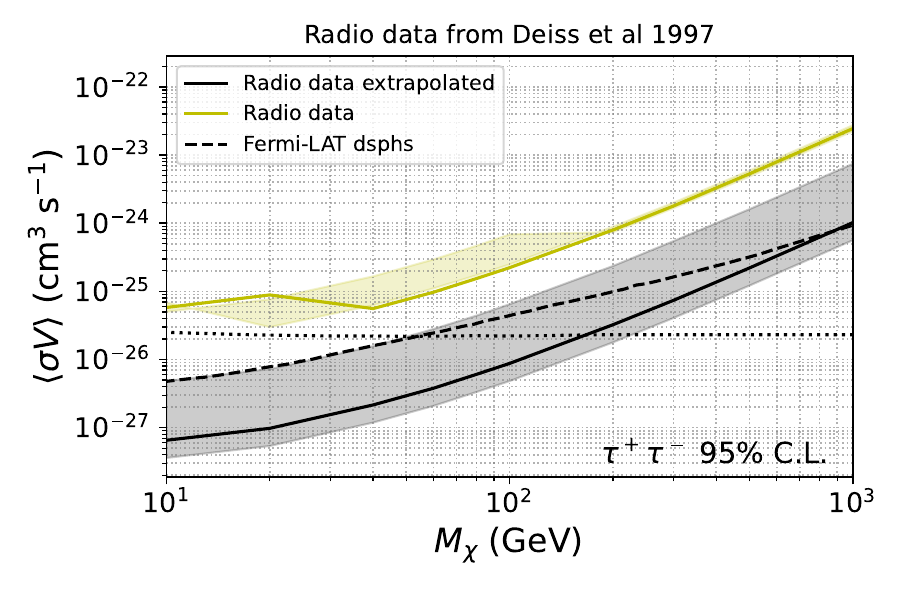}}
	\caption{95\% Confidence interval upper limits on $\langle \sigma V\rangle$, with annihilation via $\tau$-leptons, derived from the Coma cluster. The shaded bands represent the variation due to choice of halo profile. \textit{Top}: Using data set from Thierbach et al 2002~\cite{thierbach2002}. \textit{Bottom}: Deiss et al 1997~\cite{deiss1997}.}
	\label{fig:adi-tautau}
\end{figure}

All the above results were computed using The Green's function method. Here we present results derived via ADI solution. These should be more accurate, in the sense they have spatial resolution not afforded to the Green's case. This is especially important when using surface brightness data, as it is inherently spatially resolved. The results are displayed in Figures~\ref{fig:adi-bb}, \ref{fig:adi-mumu}, and \ref{fig:adi-tautau}. Interestingly, the surface brightness results improve by around a factor of 1.5 in terms of the largest mass that can be ruled out at the relic level (at least for the extrapolated resolutions). This means that we have the potential to exceed the recent limits produced via ASKAP observations of the large magellanic cloud~\cite{Regis_2021}. Conversely, the leptonic channels actually weaken in the integrated flux case, with $b\bar{b}$ being slightly stronger in contrast. This suggests that the more accurate solution method backs up the Green's method results and amplifies the potential of surface brightness probes at high resolution.   

\section{Discussion \& Conclusions}
\label{sec:disc}

In this work we have examined the potential of high angular resolution radio studies of galaxy clusters as probes of DM. This was done via the extrapolation of known surface brightness profiles down to 10$^{\prime\prime}$, a conservative estimate for what is attainable with the MeerKAT instrument. This extrapolation is justified by the fact that Ophiuchus and Coma have radio halos that show no small scale features down to $\sim 24$ arcseconds~\cite{murgia2009,govoni2010} and $1$ arcminute~\cite{deiss1997,brown-rudnick-2011,coma2022} respectively. Even with the known diffuse backgrounds in our targets, we managed to show that competitive limits can be produced with Einasto and NFW halo geometries, as their density profiles result in surface brightnesses that clash with the exponential profile of existing emissions. This is important, as there is some literature evidence that favours these geometries generally in galaxy clusters~\cite{mamon2019b,He_2020}. Note, however, that neither Coma nor Ophiuchus show any particular evidence favouring either cores or cusps~\cite{coma-halo-2003,durret2015}. 

Notably, our Green's function results, using an NFW-like Einasto profile ($\alpha_e = 0.17$), in the Coma cluster, at $1.4$ GHz, are competitive with the strongest existing indirect limits in the literature~\cite{Regis_2021}, as we potentially rule out WIMPs annihilating via $b$-quarks with $M_\chi \lesssim 400$ GeV. In so doing, we can exceed Fermi-LAT limits by around a factor of 4-5. Even in the pessimistic case of a shallowly cusped profile, our limits better the Fermi-LAT results from dwarf galaxies~\cite{Hoof_2020}. An actual NFW profile produces weaker constraints than Einasto, unless the surface brightness is extrapolated down to 1$^{\prime\prime}$ (due to differences in the profiles when $r > 10^{-3} r_s$). When the spatially resolved ADI method is used to solve for electron distributions, we are able to improve the Coma extrapolation limits by around a factor of 1.5, ruling out annihilation via $b$ quarks for $M_\chi \lesssim 700$ GeV. This demonstrates the robustness of our results. In addition to this, we determined that if Bohmian, rather than Kolmogorov, diffusion is employed, the limits do not vary significantly on the scales studied. Suggesting that variations due to diffusive assumptions are relatively small compared to differences induced by the choice of halo profile.

Importantly, our surface brightness results in Coma are superior to those derived from integrated fluxes, whose competitiveness depends upon uncertain substructure boosting effects. The results from Ophiuchus tend to be weaker than Coma, largely due to the lower concentration parameter of the DM halo. Despite this, the results exceed Fermi-LAT and and ATCA limits from dwarf galaxies~\cite{Hoof_2020,Regis_2014,regis2017}. Additionally, the results from Ophiuchus have one point to recommend them: the extrapolation down to 10$^{\prime\prime}$ is less drastic than from the existing Coma cluster data. This concern can be mitigated by a lack of observed small-scale spectral structure in LOFAR studies of Coma~\cite{coma2022}, indicating that the extrapolation may not be problematic.

These results indicate that examining galaxy clusters at arcsecond resolutions can be a highly effective tool for probing DM. Notably, we chose Coma and Ophiuchus, with fainter emissions relative to their virial mass than several other characterised radio halos~\cite{murgia2009}. This means that newer telescopes like MeerKAT may be able to produce powerful limits on DM even in the re-examination of clusters with known halos. Especially since the best limits were found around $1$ GHz, where MeerKAT is most sensitive. These galaxy cluster targets have one significant advantage over dwarf galaxies: the magnetic fields and diffusive environments are far less uncertain. Thus, even with the larger baryonic backgrounds, galaxy cluster probes can act as a powerful complement to dwarf galaxy searches by providing robust constraints, which can also be surprisingly powerful in the right environments and with the right instruments.

\bibliographystyle{apsrev4-2}
\bibliography{high_res_dm.bib}

\end{document}